# Nanoassembly of quantum emitters in hexagonal boron nitride and gold nanospheres

*Minh Nguyen†, Sejeong Kim†, Toan Trong Tran, Zai-Quan Xu, Mehran Kianinia, Milos Toth\*, and Igor Aharonovich\**

*School of Mathematical and Physical Sciences, University of Technology Sydney, Ultimo, New South Wales, 2007, Australia*

*† These authors contributed equally to this work*

[*milos.toth@uts.edu.au](*milos.toth@uts.edu.au)

[*igor.aharonovich@uts.edu.au](*igor.aharonovich@uts.edu.au)

ABSTRACT: Assembly of quantum nanophotonic systems with plasmonic resonators are important for fundamental studies of single photon sources as well as for on-chip information processing. In this work, we demonstrate controllable nanoassembly of gold nanospheres with ultra-bright quantum emitters in 2D layered hexagonal boron nitride (hBN). We utilize an atomic force microscope (AFM) tip to precisely position gold nanospheres to close proximity of the quantum emitters and observe the resulting emission enhancement and fluorescence lifetime reduction. A fluorescence enhancement of over 300% is achieved experimentally for quantum emitters in hBN, with a radiative quantum efficiency of up to 40% and a saturated count rate in excess of $5\times10^6$ counts/s. Our results are promising for future employment of quantum emitters in hBN for integrated nanophotonic devices and plasmonic based nanosensors.

Hybridization of quantum emitters and photonic constituents is highly sought after to control the nature of quantum light and the realization of integrated quantum photonic circuits.[1-9] Key requirements for these hybrid systems include robust single photon emitters (SPEs) which are bright and optically stable, as well as photonic resonators that can be readily engineered. SPEs in layered hBN are emerging sources with great potential for a variety of quantum applications due to their high brightness, long-term stability and full in-plane polarization.[10-15] It is therefore interesting to further study the enhancement of these SPEs *via* the Purcell effect. Achieving such an enhancement, however, requires maximal spectral overlap between the cavity resonance and the source emission as well as spatial positioning of the emitter in the maximum of the cavity electromagnetic field.

Plasmonic nanoantennas are advantageous among a variety of optical resonators due to their broad spectral resonances, which make spectral matching between the optical mode and an emitter substantially less complicated.[3, 16-18] Although plasmonic resonances have low Q-factors compared with dielectric cavities, they provide high Purcell enhancements due to small mode volumes,[3, 19] also known as 'hot spots'. Several techniques including electron beam lithography,[20] optical trapping of quantum emitters on top of metal nanoantennas[21] and wet chemical transfer[17] have been previously explored for deterministic assembly of hybrid quantum nanophotonic–plasmonic devices. While this tight light confinement makes them ideal components for optical integrated chips,[22-24] challenges still remain with the establishment of reliable methods to arrange quantum emitters within the hot spots of plasmonic cavities. An atomic force microscope (AFM) provides a promising solution to undertake this challenge by virtue of its capability in manipulation of nanosized objects with nanometer precision, combined with a simultaneous high-resolution imaging capability.[1, 19, 25]

In this work, we deterministically assemble and study a hybrid plasmonic-quantum system comprised of a SPE in layered hBN coupled with gold nanospheres. Single and double-plasmonic particle arrangements were realized experimentally by placing gold nanospheres precisely in close proximity to a pre-characterized SPE in hBN using an AFM tip. Coupling the SPEs to gold nanospheres modifies the spontaneous emission and enables an ultra-bright source of non-classical light. Reduced fluorescence lifetime accompanied by a photoluminescence (PL) enhancement is observed in power-dependent PL measurements. Our results help to highlight the potential of plasmonic enhancement of SPEs hosted by 2D layered materials for future applications in photonic integrated chip technology.

A schematic of an optically-active defect, namely the antisite nitrogen vacancy ($N_BV_N$), in a 2D hBN lattice is shown in Figure 1a. The size of the hBN flake is crucial as it determines the possible distance between the quantum emitter and the gold nanosphere. Smaller sizes of flakes increase the probability of positioning quantum emitters within plasmonic hot spots. We therefore chose emitters which are embedded in isolated, small flakes for this study. Figure 1b shows an AFM image of a typical hBN flake with in-plane dimensions of approximately 560 nm in x and 440 nm in y, at their respective widest points. The height of the flake is similar to that of the nanosphere, with a vertical thickness of ~ 84 nm (See supporting information figure S1). The inset of Figure 1b is a confocal PL image taken from the same flake revealing the bright emission corresponding to the $N_BV_N$ defect.

The optical characteristics of an emitter embedded in the aforementioned flake are shown in Figure 1c. A room temperature PL spectrum of the emitter reveals a sharp zero phonon line (ZPL) at 578 nm which is in the typical range for SPEs in hBN,[10] with a full-width at half-maximum (FWHM) of 2.33 ± 0.04 nm. The phonon side-band (PSB) is also visible at 635 nm and is spectrally separated from the ZPL by over 40 nm from peak to peak. The Debye-Waller (DW) factor for this emitter is 0.75, as calculated from the ratio of ZPL emission to total emission. For the measurements throughout the manuscript, we performed spectral filtering to further isolate the ZPL and reduce transmission of signal from other background sources with a 20 nm tunable optical bandpass filter. The luminescent centre is shown to produce antibunched emission characterised by its second-order correlation function, exhibiting a $g^{(2)}(0)$ value of 0.27 (Figure 1c, inset). The deviation from zero is attributed to the residual background and the detector jitter. The dip below 0.5 is a clear signature for single-photon emission. The linearly polarized emission transition dipole of the SPE was measured and fit to a cosine function, $cos^2(\theta)$, as shown in Figure 1d. The emission is linearly polarized at an angle of 67° with respect to the y-axis. The y-axis is aligned at 0° and the x-axis is aligned at 90° (Figure 1b, inset). The degree of polarization, defined as $\frac{I_{max}-I_{min}}{I_{max}+I_{min}}$ (where $I_{max}$ and $I_{min}$ are maximum and minimum intensities, respectively), is 0.83. Such a high degree of polarization is advantageous for applications in quantum technologies.

We start by investigating the theoretical fluorescence enhancement caused by positioning a single and a double gold nanosphere in a proximity of the emitter using the finite difference time domain (FDTD) method. Fluorescence enhancement originates from both excitation enhancement ($\gamma_{exc}/\gamma_{exc}^0$) and spontaneous emission enhancement ($\gamma_{sp}/\gamma_{sp}^0$). Initially, we separately simulate the local enhancement of the excitation laser field. A 532 nm plane wave with x-polarized light excites localized surface plasmons on a gold nanosphere with a diameter of 50 nm. Note that 50 nm gold nanospheres are used both in simulation and experiments, and its absorption cross-section overlaps well with the emission spectrum of the hBN SPE. (See supporting information Figure S2) The local excitation enhancement by the gold nanosphere versus distance from the particle is plotted in Figure 2a. Because localized plasmonic fields are most intense near the metal surface, as shown in the optical field intensity profile in the inset of Figure 2a, the local excitation enhancement decreases with increasing distance from the metal surface. Aside from the excitation enhancement, the radiative decay rate of the quantum emitter is also enhanced by the Purcell effect. The spontaneous emission rate enhancement caused by a gold nanosphere ($\gamma_{sp}$) relative to the emission rate in free space ($\gamma_{sp}^0$) is shown in Figure 2b. Here, the emitter dipole is assumed to have an emission peak of 580 nm, which matches the experimentally studied hBN quantum emitter. The hBN refractive index is assumed as $n_x=n_y=1.84$ and $n_z=1.72$ for the purpose of the simulation. Quantum emitters which are x-polarized, perpendicular to the metal surface, show higher enhancement than those that are y-polarized. Under ideal conditions, an enhancement of more than 520 and 330 times is expected for x-polarized and y-polarized dipoles, respectively. The radiative enhancement factor, defined as the power radiated into free space in the presence of the metal sphere ($W_r$) divided by that in the absence of the sphere ($W_0$), is plotted in Figure 2c. In addition, we calculated the spontaneous emission rate enhancement versus the gap distance between two gold nanospheres (Figure 2d). The double gold nanosphere arrangement creates a hot spot in the gap, where the intensity is determined by the gap size. In this configuration, Purcell enhancement of more than 1000 times were expected under optimized conditions.[5] As the metal gap size increases, the enhancement drastically decreases, converging to 1 for the dipole to metal surface greater than 60 nm, corresponding to a null gap plasmon effect on the dipole source.

To investigate the experimental enhancement, an AFM (Dimension 3100) was used to deterministically couple gold nanospheres to pre-characterized hBN quantum emitters. Gold nanospheres were physically manipulated by direct contact with the AFM tip and moved in small increments towards the hBN flake shown in Figure 1, as is illustrated in Figure 3a. Initially, a single gold nanosphere is moved into position to couple a plasmonic cavity to the SPE embedded within the flake (Figure 3b,c). This particular arrangement is denoted as the 'single particle' configuration. A second gold sphere was then coupled to the hBN flake in a similar fashion (Figure 3d-f). Similarly to the 'single particle' configuration, movement of the second particle into position with the AFM can be seen in Figure 3(d-f). We refer to the resulting arrangement as the 'double particle' configuration. The second nanosphere is moved to align with the transition dipole angle shown in Figure 1d in order to maximize the plasmonic coupling effect as predicted with the simulation results in Figure 2. This transition dipole angle was deduced from the orientation of the flake relative to nearby reference markers. We also note that despite the second particle not completely aligning with the first particle, it is still positioned within the maximum emission field when cross-referenced from both the AFM image and the polarisation measurement.

We performed detailed optical characterization of each configuration (pristine SPE, "single" and "double" particle) using a home-built confocal microscope and a Hanbury-Brown and Twiss (HBT) interferometer setup (see supporting information Figure S3). Time-resolved PL from this hBN SPE shows a reduction in emission lifetime from $\tau_{pristine} = 4.92$ ns to $\tau_{single} = 2.68$ ns, for the single-particle arrangement, and consequently, a further emission decay to $\tau_{double} = 1.54$ ns for the double-particle arrangement. The data are plotted in Figure 4a and Figure 4b, showing spontaneous emission rate enhancement factors of 1.84 and 3.19 for the single- and double-particle arrangements, respectively. The decrease in emission lifetime is attributed to tight confinement of the electric field, resulting in an increase of the local density of states (LDOS). The measured lifetime is a total comprised of the radiative and non-radiative components: $1/\tau_{tot} = 1/\tau_r + 1/\tau_{nr}$. Therefore, to confirm that the lifetime reduction seen in Figure 4a and 4b is caused primarily by the increase in the radiative decay channels, the saturation intensities were determined from the PL versus laser power curves shown in Figure 4c. The saturation measurements were fitted with the following equation:

$$I = \frac{I_\infty P}{(P_{Sat} + P)}$$

where $I_\infty$ is the saturated count rate and $P_{Sat}$ is the saturation power. The saturated count rates are $2.89 \times 10^6$ counts/s and $5.79 \times 10^6$ counts/s for the single- and double-particle arrangements, respectively. This translates to overall enhancement factors of 1.55 and 3.10 relative to the uncoupled, pristine hBN emitter which exhibits a saturated intensity of $1.87\times10^6$ counts/s. A comparison of the measured decay rate enhancement with the saturated intensity enhancement reveals that the latter originates primarily from enhancement of the radiative decay channel. Remarkably, the hybrid quantum plasmonic system generates almost ~ $6\times10^6$ counts/s, representing one of the brightest room temperature SPE in a layered material reported to date, associated with a modest Purcell enhancement of 3.

To confirm that the quantum nature of the emitter is preserved, we compared the second-order autocorrelation function, ($g^{(2)}(\tau)$), for the three configurations. Figure 4d shows the plots of the $g^{(2)}(\tau)$, offset vertically by 0.5 for clarity. The $g^{(2)}(0)$ value across the three data sets is consistent with minimal changes within the uncertainty of the measurements, with values of $g^{(2)}(0)=0.24$, $g^{(2)}(0)=0.26$, $g^{(2)}(0)=0.31$ for the pristine, single, and double particle arrangements, respectively.

For further analysis of the hybrid system, PL enhancement in the un-saturated regime can be approximated by the following term:

$$EF \propto \eta \times \gamma_{exc} \times QE$$

where $\eta$ is the collection efficiency, $\gamma_{exc}$ is the excitation rate, and QE is the quantum efficiency of the emitter in the coupled system.[5] For the studied configurations, we expect changes in collection efficiency to be negligible with a high NA objective lens (NA=0.9) for the pristine and the coupled emitter configurations.[2] Below the saturation threshold, the net enhancement has an excitation contribution caused by strong localization of fields around the metal nanosphere which serves to increase absorption of the excitation laser by the coupled system. This typically manifests as an increase in the slope of the saturation curve.[7] We note that emitter count rates measured beneath saturation, at 300 μW, are higher than the saturated enhancement factors with enhancement factors of 1.69 for the single particle and 4.18 for the double particle arrangements. We attribute this difference to increasing excitation enhancement onset by an increased absorption cross-section upon the introduction of the nanospheres. We also note that the power, at which the system begins to saturate, decreases with the second nanosphere addition. Measurements show saturated power values of 0.88 mW, 0.78 mW, and 0.57 mW for the pristine, single particle, and double particle arrangements, respectively, supporting our claim.

On the other hand, above the saturation threshold, the emission is proportional to the radiative lifetime.

$$EF_\infty \propto \eta \times \gamma_r \times QE$$

For the case of the double sphere, $\tau_{double} = 1.54$ ns, $P_{sat} \sim 5.79 \times 10^6$ counts/s and the setup efficiency (from objective lens to the detector) of our experimental setup is estimated at ~ 2%, yielding a high radiative quantum efficiency of ~ 40%.

In summary, we demonstrated deterministic coupling of SPEs in layered hBN to plasmonic gold nanospheres. Accurate nanoscale manipulation of the gold nanospheres by an AFM tip was used to realize two hybrid coupled systems comprised of a quantum emitter in hBN, and one and two gold nanoparticles. An emission enhancement associated with a lifetime reduction was observed, yielding an impressive overall count rate from a single emitter of more than $5 \times 10^6$ counts/s at room temperature. The presented technique can be applied to a range of hybrid plasmonic–photonic systems for studying other 2D materials, such as transition metal di-chalcogenides[26-27] and exfoliated hBN monolayers. Further optimization in positioning and improvements in collection efficiency are expected to yield even higher count rates, which will be very attractive for practical devices.

**EXPERIMENTAL SECTION:**

**Sample Preparations**. Commercial solvent-exfoliated hBN crystals (Graphene Supermarket Inc.) were drop-casted onto 1 × 1 cm$^2$, silicon substrates with a 300 nm thick thermal oxide capping layer. Gold markers were deposited on the substrates using standard photolithography to locate the same flakes of choice easily. hBN crystals were then annealed at 850 °C in quartz tube furnace fluxing Ar gas under 1 torr for 30 minutes to activate the emitter. Colloidal 50 nm gold nanospheres (nanoComposix) in stock solution were centrifuged multiple times and redispersed into MilliQ water to ensure the nanosphere solution was free from large foreign particulates.

**Simulated Modelling of Plasmonic Response**. Numerical calculations were performed with finite difference time domain (FDTD) methods. Here, we used the refractive index determined by Johnson and Christy[28] for gold nanoparticles. For birefringent hexagonal boron nitride, we extrapolated two refractive indices, ordinary and extraordinary refractive indices, from the experimental data in the infrared region. The whole computational domain was divided by 1 nm spatial grids.

**Particle movement and Characterisations.** A Veeco Dimension 3100 AFM from Digital Instruments equipped with a NSC35 silicon cantilever (spring constant of k = 4.5) was used for topological imaging in tapping mode and nanomanipulation in contact mode. Confocal PL scanning and PL spectra were performed under a 532 nm continuous wave (CW) laser (Gem 532, Laser Quantum Ltd.) excitation. The laser beam was passed through a half waveplate and focused upon the sample *via* an air objective lens (NA = 0.9, Nikon). X-Y scanning was performed with a Newport FSM-300 piezo scanning mirror. Collected light was filtered via a 532 nm dichroic mirror (532 nm laser BrightLine, Semrock), with further spectral filtering performed with a tuneable 20 nm bandpass filter. The emission from the sample is collected and directed into a graded index fiber, which serves as the confocal pinhole. The signal was directed towards a flip-mirror which can be controlled to guide the emission towards either a HBT interferometer for photon correlation measurements or a spectrometer (Acton SpectraPro, Princeton Instrument Inc.) for spectra collection. The HBT interferometer consists of two avalanche photodiodes (APD, Excelitas Technologies), with a 100 ns delay time induced in one of the APDs connected to a time-correlation card (PicoHarp 300). Time-resolved PL measurements were performed with a 512 nm pulsed excitation laser (PiL051X, Advanced Laser Diode Systems GmbH), with a 100 ps pulse width and a 10 MHz repetition rate.

# FIGURES:

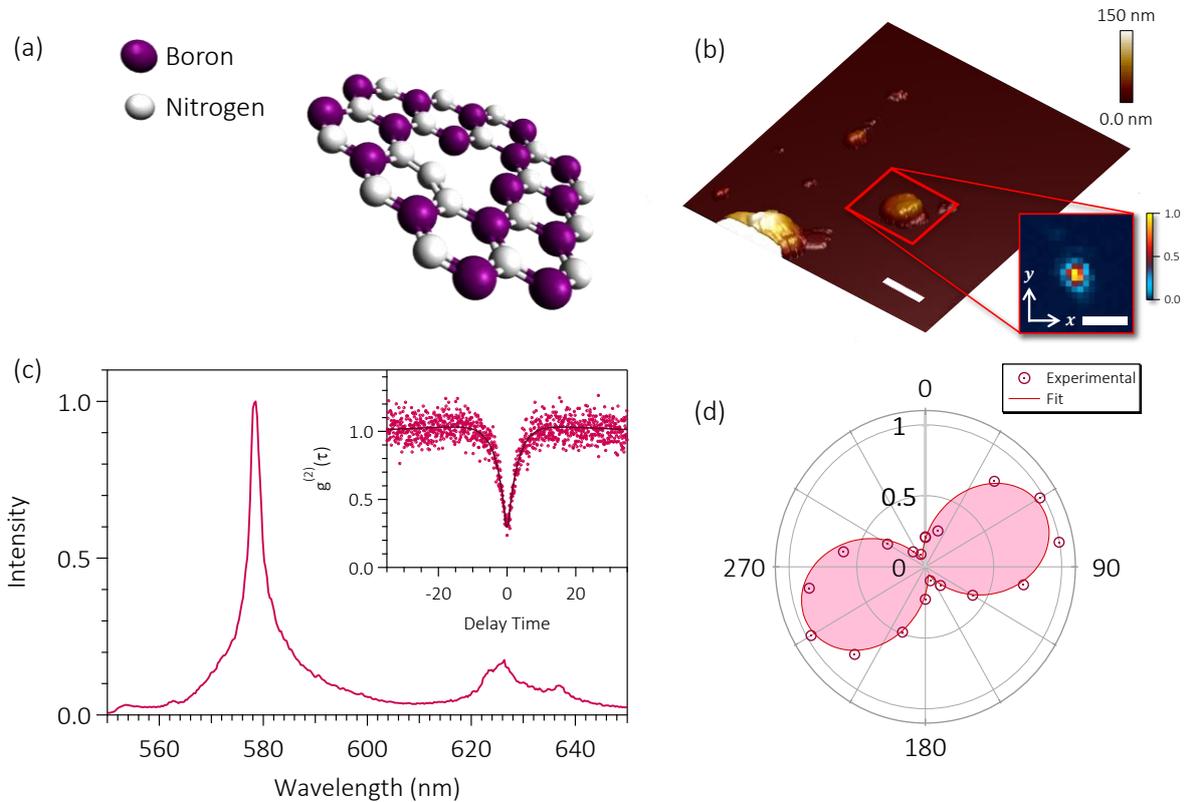

**Figure 1.** Optical characteristics of the luminescent hBN defect. (a) Schematic illustration of the nitrogen-antisite vacancy centre, $N_BV_N$. (b) AFM image showing the morphology of the hBN flake used for plasmonic coupling. The flake dimensions span approximately 560 nm in x and 440 nm in y, at their respective widest points. Inset: confocal PL image of the hBN SPE. The scale bar is 1µm (c) Fluorescence spectrum of the hBN SPE. The emitter ZPL is at 578 nm with a full-width at half-maximum (FWHM) of 2.33 ± 0.04 nm. Inset shows the second-order autocorrelation function with $g^{(2)}(0) = 0.24$, indicating single photon emission. (d) Polar graph of emission showing linearly polarized emission. The linear transition dipole is orientated at 67° with respect to the north pole of the flake with an orientation as observed in (a) inset. All optical measurements are performed with 532 nm laser at room temperature.

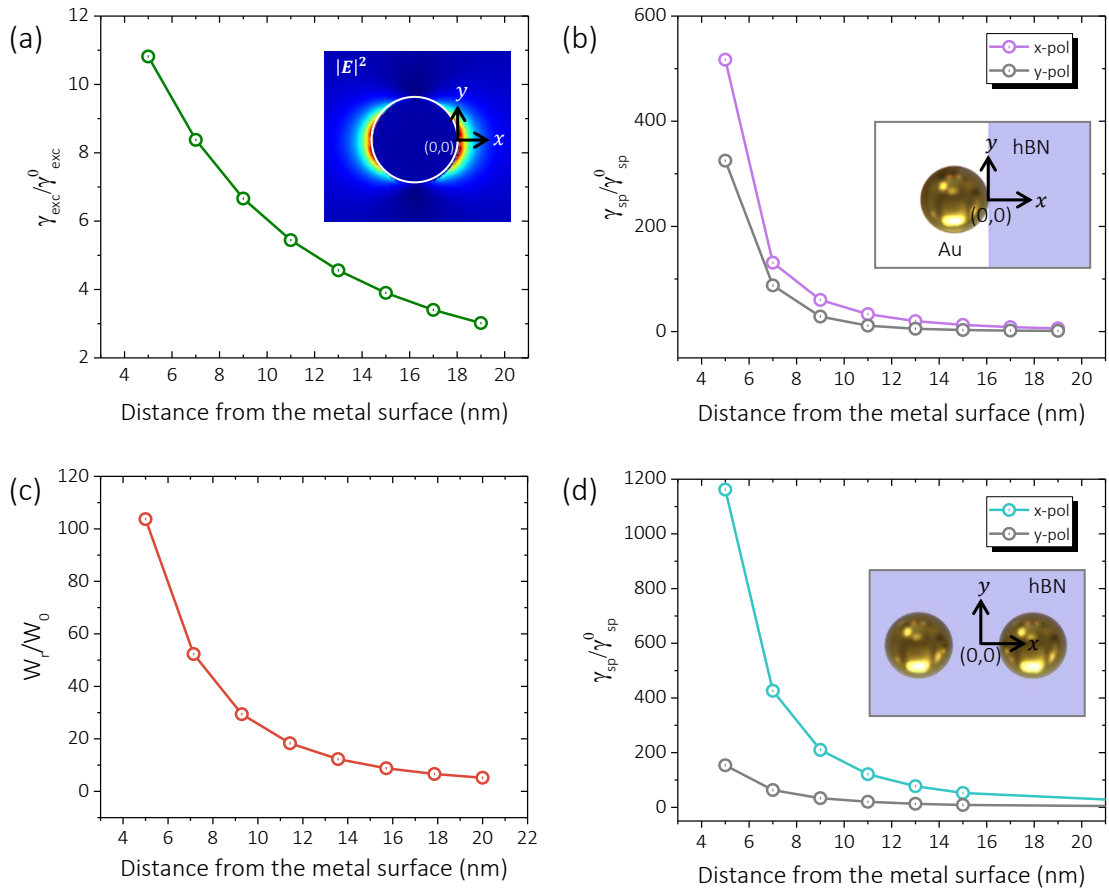

**Figure 2.** FDTD simulations of the optical response of two distinct coupling arrangements. (a) Excitation enhancement as a function of distance from a nanosphere surface. Inset: an FDTD simulation of electric field intensity of a gold nanosphere excited by x-polarised light. (b) Purcell factor as a function of distance from the emitter dipole. The purple and grey traces represent simulated Purcell factors for x- and y-polarized emitters, respectively. (c) Radiated power enhancement as a function of distance between the nanosphere and dipole. (d) Purcell factor for both x (teal trace) and y-polarized (grey trace) light as a function of the gap distance between two nanospheres. The emitter dipole is assumed to be at the origin and is equidistant from both nanospheres. The gap plasmon effect on the dipole source from the two particles diminishes drastically as the metal gap size increases.

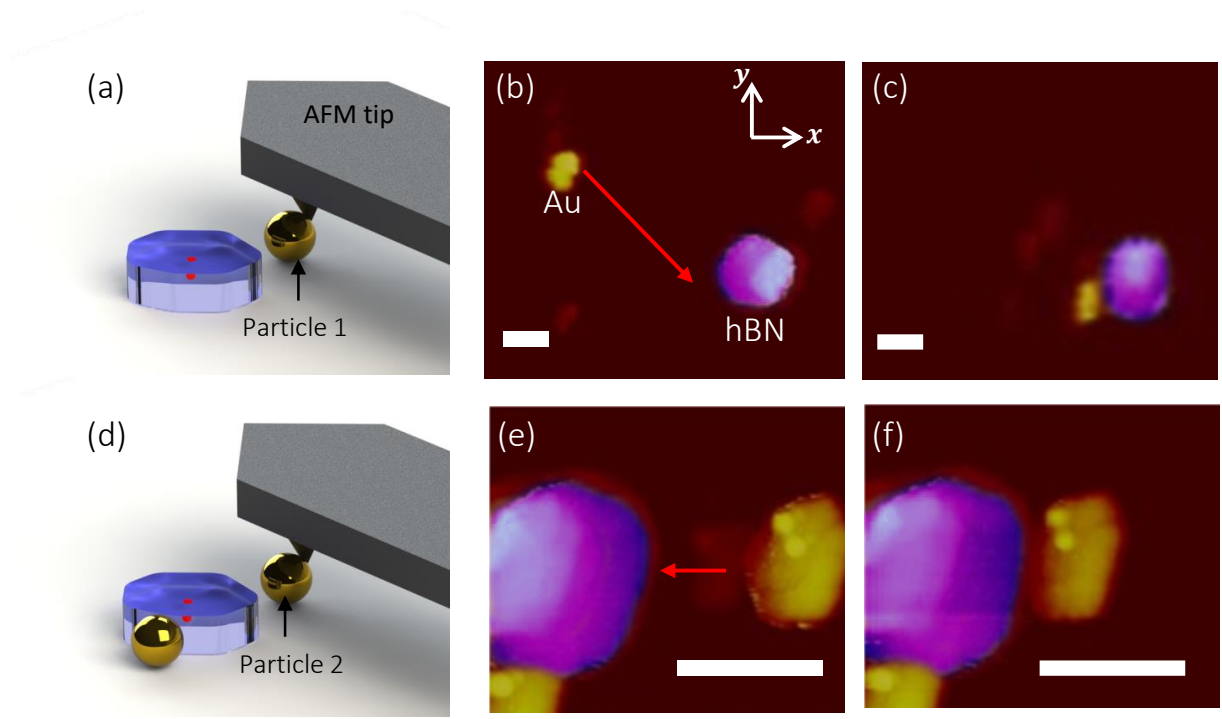

**Figure 3**. Deterministic nanomanipulation of gold spheres for plasmonic coupling using an AFM tip. (a) Schematic illustration of the movement of a gold sphere towards an SPE in hBN using an AFM tip. Movement of the gold sphere is induced from physical contact between the tip and the particle. The luminescent defect (red) is embedded within the hBN flake (purple). (b) Initial positioning of the gold sphere in close proximity to the hBN flake hosting the SPE. The red arrow indicates the direction of movement. (c) The final arrangement of the gold sphere and the hBN nanoflake after positioning. The particle is in contact with the flake. (d) Schematic illustration of the movement of the second gold sphere to the same hBN flake. (e) The second sphere (on the right) and its movement direction shown by the red arrow. The orientation of the image has been shifted counter-clockwise (with reference from (b)-(c)) by several degrees in order to align the particle for movement towards the flake. (f) The final position of the second particle in contact with the flake. All scale bars are 250 nm. The hBN flake was false color coded for clarity.

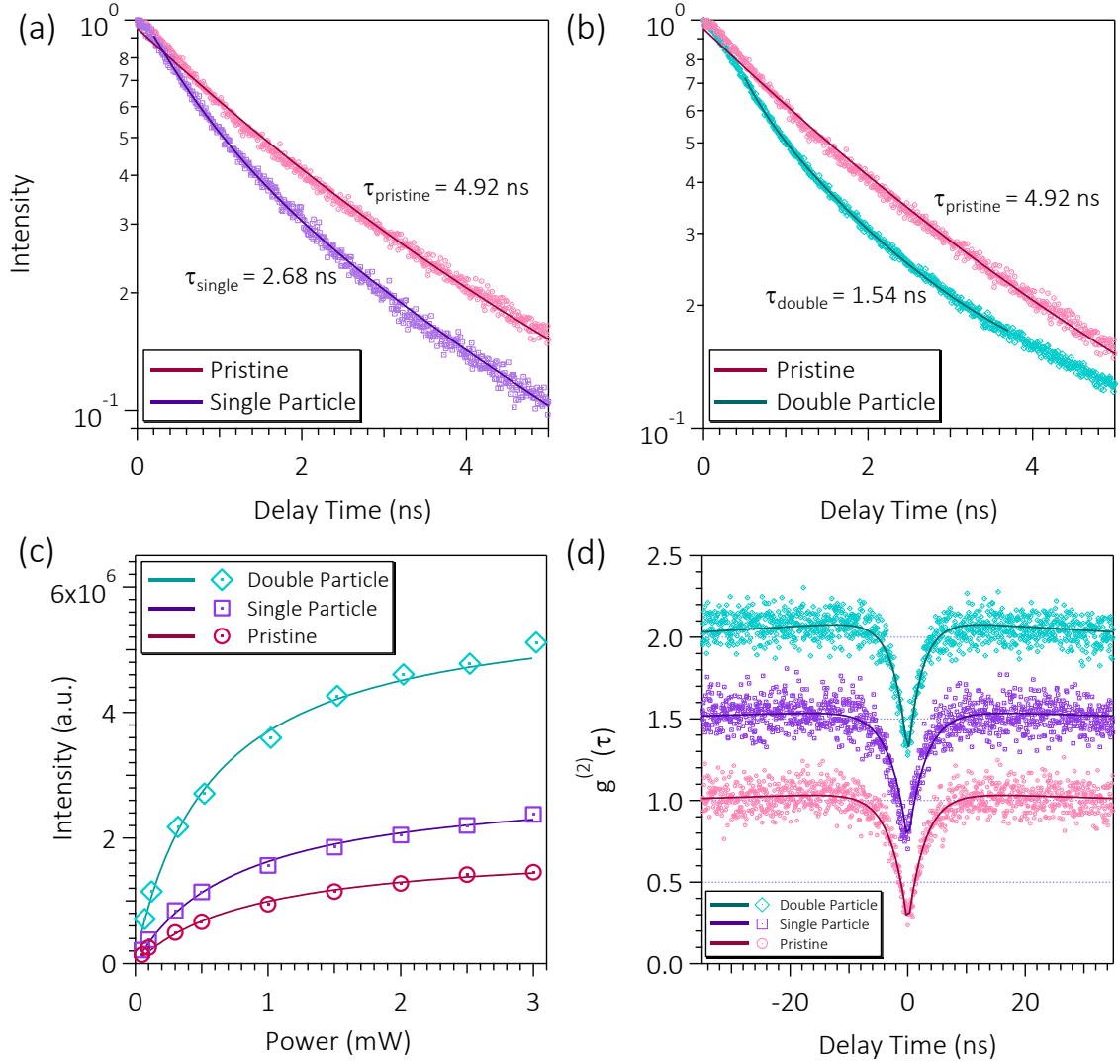

**Figure 4**. Characterisation of the optical response from plasmonic coupling. Data from a pristine emitter, a single particle arrangement, and a double particle arrangement are shown in pink, purple and teal, respectively. (a) A comparison between time-resolved PL measurements of pristine and single particle configurations, and (b) double particle arrangement. (c) A comparison of fluorescence saturation curves between the pristine, single particle, and double particle arrangements. (d) Second-order autocorrelation functions for the three configurations. All $g^{(2)}(0)$ values are below 0.5, confirming single photon emission. All measurements were recorded at room temperature.

AUTHOR INFORMATION

**Corresponding Authors**

*E-mail: igor.aharonovich@uts.edu.au. *E-mail: milos.toth@uts.edu.au.

ACKNOWLEDGEMENTS

The authors would like to acknowledge the Financial support from the Australian Research Council (DE130100592), FEI Company, the Asian Office of Aerospace Research and Development grant FA2386-15-1-4044 are gratefully acknowledged. This research is supported in part by an Australian Government Research Training Program (RTP) Scholarship.